# A Recursive Delay Estimation Algorithm for Linear Multivariable Systems with Time-varying Delays

Authors: *Iman Shafikhani* [1], *Hazhar Sufi Karimi* [2], *Mohammad Mohammadian* [3], *Amin Ramezani* [4], *Hamid Reza Momeni 4*

Abstract: Time delay estimation plays a critical role in control, stabilization and state estimation of many practical system with time delay. In this paper, we propose a method to estimate delay for discrete time linear multiple-input multiple-output systems with time-varying input delays. This method is purposefully given for situations where only a limited amount of information is available for the system. Although, this approach is primarily developed in a deterministic framework, it can also be applied to noisy data under special circumstances. In addition, switched linear autoregressive models with exogenous inputs are introduced as possible applications of the presented algorithm provided that the switching frequencies are small. Finally, effectiveness of the algorithm is illustrated by two numerical examples.

## 1. INTRODUCTION

Time delay systems have attracted a lot of attention due to their prevalence in various practical areas, such as networked control systems, biology, economy, chemical processes, neural networks, etc. [1], [2], [3]. Accordingly, many promising methodologies (for control, stabilization, estimation, etc.) have been developed to deal with this important class of systems [4]. Most of these methods are model-based, and hence, their success critically depends on accuracy of the system models being used. Actually, performance of model-based methods deteriorates if they are designed according to inaccurate models, regardless of other factors. This fact demonstrates the importance of developing effective system identification methods for time delay systems.

Identification of these systems, which happen to be multivariable in many cases, is usually done in two different ways. One approach is to estimate time delays by available methods and then shift data in time according to delay estimates to identify delay-free dynamics [5-8]. The other approach is to estimate model parameters and time delays simultaneously ([9-12]). Identification of continuous time delay systems has been addressed in [13]. In [14], Nonlinear time delay systems has been identified by using wiener approach. It should be noted that delays may be time varying due to aging or as intrinsic properties of systems. Furthermore, identification techniques are usually implemented online for some practical purposes such as real-time applications and adaptive control. Hence, it is desirable to estimate time delays recursively. Quite a few methods have been proposed to estimate time-varying delays [6], [7], [15-17]. Most of these techniques require prior knowledge about system structure. For example, many of them assume that the degrees of numerator and denominator polynomials in transfer function of system is known in advance (see [9], [16] and [18] for instance). Such requirements are definitely restrictive when there is a lack of needed information. This issue is more evident in dealing with multivariable systems where even a small amount of missing information can lead a method to failure. It is therefore essential to have an alternative approach which recursively estimates time delays while relying on the least prior knowledge possible. Two recent contributions are worth mentioning in this regards; [19] employs a commonly used method in signal processing called compressed sensing (CS) recovery theory, and formulates delay/parameter identification as a CS recovery problem. In addition, the state-space realization approach presented by [20] utilizes combination of input pulse signals to explicitly determine time delays. Both of these methods are advantageous in that they perform estimation without the need for restrictive prior information. However, they are essentially batch methods suggesting that they are not suitable for online implementation due to computational issues. This problem is the main motivation of the present study, and will be addressed using subspace framework. In fact, our goal is to develop an online method for delay estimation which does not require information about system structure.

[1] Vehicular Systems , Department of Electrical Engineering, Linköping University, Sweden
2 Department of Electrical and Computer Engineering, Kansas State University, USA
3 Department of Electrical Engineering, Arak University of Technology, Iran
4 Electrical Engineering Department, Tarbiat Modares University, Tehran, Iran

The main reason for choosing subspace framework is that it has proven to be successful in dealing with black-box modeling problems [21]. It is due to the fact that subspace methods are capable of relying solely on the input-output measurements. This merit, however, does not limit one to incorporate prior information into system identification ([22],[23]). Furthermore, they are easily applicable to MIMO systems [24]. In fact, these methods are developed for state-space models, and in this way they do not discriminate between MIMO and SISO models. This implies that subspace methods do not suffer from parameterization issues which are troublesome for multivariate system identification (Note that no single linear parameterization can cover all linear MIMO models [25]). They are also very useful for control synthesis as they have been integrated for closed-loop identification [26]. As another advantage, subspace methods utilize numerically robust techniques such as QR factorization and singular value decomposition, which make them even more attractive [21].

In this paper, we look at subspace identification as a tool for delay estimation. The proposed algorithm is based on the key idea that time delay is the time needed for the impulse response to start [27]. Once Markov parameters of system (which are the impulse responses of it) are estimated, time delays can be determined implicitly. This idea was also used in [28] using a subspace method. However, our idea is to develop a recursive algorithm for estimation of Markov parameters by which the computational cost is significantly reduced. We use the same deterministic framework as [28] did, and we show that this method is usable for noisy data if some conditions are satisfied. As an important application, the proposed algorithm is applicable to SARX models with time delays for small switching frequencies. This is shown by a numerical example that validates the theoretical results.

The rest of the article is organized as follows: Section 2 is dedicated to problem formulation with a brief review of the requisite background; in Section 3, we give our algorithm and the relevant explanations; two numerical examples in Section 4 assess the effectiveness of the proposed algorithm, and ultimately a conclusion is provided in Section 5.

## 2. PROBLEM STATEMENT

Consider the discrete time MIMO model

$$
\begin{cases}
y_1(k) = \sum_{r=1}^{m} G_{1r}(q^{-1}) u_r(k-1-T_{1r}(k)) \\
y_2(k) = \sum_{r=1}^{m} G_{2r}(q^{-1}) u_r(k-1-T_{2r}(k)) \\
\vdots \\
y_l(k) = \sum_{r=1}^{m} G_{lr}(q^{-1}) u_r(k-1-T_{lr}(k))
\end{cases}
\quad \textbf{(1)}
$$

in which $u_i(i = 1,2,\dots,m)$ is the $i$th input, $y_j(j = 1,2,\dots,l)$ is the $j$th output, $G_{ji}$s are strictly proper, stable transfer functions, and $T_{ji}$s are time delays that we aim to estimate them in this paper. It is desirable for the method to be recursive so that one can track changes in time delays. In addition, the method is intended to work successfully regardless of having any information about $G_{ji}$s (like their numerators' or denominators' degrees).

Before giving our method, we briefly explain a recently proposed subspace-based algorithm for estimation of time delays [28]. The main advantage of this algorithm is that it does not need any information about the model's structure. To show how the algorithm works, consider a state-space form of a deterministic linear time invariant system with $m$ inputs and $l$ outputs

$$
\boldsymbol{U_p} \triangleq \boldsymbol{U_{0,i,j}} = \begin{pmatrix} \boldsymbol{u_0} & \boldsymbol{u_1} & \dots & \boldsymbol{u_{j-1}} \\ \boldsymbol{u_1} & \boldsymbol{u_2} & \dots & \boldsymbol{u_j} \\ \vdots & \vdots & \ddots & \vdots \\ \boldsymbol{u_{i-1}} & \boldsymbol{u_i} & \dots & \boldsymbol{u_{i+j-2}} \end{pmatrix} \quad \textbf{(2)}
$$

$$
\boldsymbol{U_f} \triangleq \boldsymbol{U_{i,h,j}} = \begin{pmatrix} \boldsymbol{u_i} & \boldsymbol{u_{i+1}} & \dots & \boldsymbol{u_{i+j-1}} \\ \boldsymbol{u_{i+1}} & \boldsymbol{u_{i+2}} & \dots & \boldsymbol{u_{i+j}} \\ \vdots & \vdots & \ddots & \vdots \\ \boldsymbol{u_{i+h-1}} & \boldsymbol{u_{i+h}} & \dots & \boldsymbol{u_{i+h+j-2}} \end{pmatrix} \quad \textbf{(3)}
$$

The same definitions go for $Y_p$ and $Y_f$. In (3) and (4), "p" and "f" represent past and future respectively. The extended observability matrix $\Gamma_k$and the extended controllability matrix$\Delta_k$are defined as

$$
\boldsymbol{\Gamma}_k \triangleq \begin{pmatrix} C \\ CA \\ \vdots \\ CA^{k-1} \end{pmatrix} \in \mathbb{R}^{kl\times n} \text{ and } \boldsymbol{\Delta}_k \triangleq \begin{pmatrix} A^{k-1}B & A^{k-2}B & \cdots & B \end{pmatrix} \in \mathbb{R}^{n\times km}.
$$

A Toeplitz matrix comprised of Markov parameters is also shown below.

$$
\boldsymbol{\Psi}_k \triangleq \begin{pmatrix} D & 0 & 0 & \cdots & 0 \\ CB & D & 0 & \cdots & 0 \\ CAB & CB & D & \cdots & 0 \\ \vdots & \vdots & \vdots & \ddots & \vdots \\ CA^{k-2}B & CA^{k-3}B & CA^{k-4}B & \cdots & D \end{pmatrix} \in \mathbb{R}^{kl\times km}
$$

Iteration of (2) leads to the following extended state-space form.

$$
\begin{cases}
Y_p = \boldsymbol{\Gamma}_i X_p + \boldsymbol{\Psi}_i U_p \\
Y_f = \boldsymbol{\Gamma}_h X_f + \boldsymbol{\Psi}_h U_f \\
X_f = A^i X_p + \boldsymbol{\Delta}_i U_p
\end{cases}
$$

**(4)**

Denoting $W_p = \begin{pmatrix} U_p \\ Y_p \end{pmatrix}$, and considering the LQ factorization

$$\begin{pmatrix} U_f \\ W_p \\ Y_f \end{pmatrix} = \begin{pmatrix} L_{11} & 0 & 0 \\ L_{21} & L_{22} & 0 \\ L_{31} & L_{32} & L_{33} \end{pmatrix} \begin{pmatrix} {Q_1}^T \\ {Q_2}^T \\ {Q_3}^T \end{pmatrix} \tag{5}$$

where $L_{11} \in \mathbb{R}^{hm \times hm}$,$L_{22} \in \mathbb{R}^{i(l+m) \times i(l+m)}$,$L_{33} \in \mathbb{R}^{hl \times hl}$, one can show that the below equation holds under a special condition [25]:

$$\widehat{\boldsymbol{\Psi}}_{\boldsymbol{h}} = (L_{31} - L_{32}{L_{22}}^{\dagger} L_{21}){L_{11}}^{-1} \tag{6}$$

In (7), "†" is the Moore-Penrose pseudoinverse operator. It should be mentioned that the quality of the above estimate depends on the number of observations ($N$) as the more observations we have, the more precise the estimate in (7) becomes. Once $\widehat{\Psi}_h$ is available, we have access to estimates of Markov parameters. Since Markov parameters are in essence impulse responses of the system, and according to the fact that time delay is the time it takes for the impulse response to start, time delays can be estimated using $\widehat{\Psi}_h$. For this to be accomplished, elements in the first block column (or the first column for SISO case) of $\widehat{\Psi}_h$ corresponding to a pair of input and output are collected into a vector. Then, counting zeros before noticing a rise in the vector's entries yields the time delay estimate corresponding to that pair (delay is smaller than the number of zeros by one because of discretization). Nevertheless, the first elements are not exactly zero because of the finite data horizon $N$, but are very small figures. A comparison can then be made with a threshold (of 0.02 for example as [28] suggested) to determine "zeros". As is clear, the algorithm relies solely on the input-output data and does not need any knowledge on model structure.

If the algorithm is applied to a moving window of data whose end is at the present instant, it becomes possible to estimate time delays in a semi-online fashion as [28] has suggested. However, computation of LQ decomposition at each instant is computationally inefficient and a recursive algorithm should be developed- which is the main goal of this study. Furthermore, comparison with a threshold has some flip sides. For example, if the Markov parameters are smaller than the threshold, estimates will often be greater than their true values. In addition, this threshold is very sensitive to noise. These issues need us to revisit the criterion. Before proceeding to the next section, the following definition, theorems and assumptions should be provided.

**Definition 1.** Let $u$ be a quasi-stationary signal with covariance function$R_u(.)$.$u$ is said to be persistently exciting of order $n$ if the matrix

$$R_n = \begin{pmatrix} R_u(0) & \dots & R_u(n-1) \\ \vdots & \ddots & \vdots \\ R_u(n-1) & \cdots & R_u(0) \end{pmatrix}$$

is positive definite.

**Lemma 1.** [29]: Let $A, B = A + E \in \mathbb{C}^{m \times n}$ and $\text{rank}(B) = \text{rank}(A) = r \le min(m, n)$. Then we have
$\|B^{\dagger} - A^{\dagger}\|_F \le \|B^{\dagger}\|_2 \|A^{\dagger}\|_2 \|E\|_F$.

**Lemma 2.** [30]: Let $A, B = A + E \in \mathbb{C}^{n \times n}$, and $\text{rank}(B) = \text{rank}(A) = r$. Then we have
$\|B^{\dagger} - A^{\dagger}\|_2 \le \mu \|B^{\dagger}\|_2 \|A^{\dagger}\|_2 \|E\|_2$,

where $\mu$ can be found in Table I.

**Table 1. Values of $\mu$ for spectral norm**

| Rank | $\mu$ |
|---|---|
| $r < n$ | $\frac{1+\sqrt{5}}{2}$ |
| $r = n$ | 1 |

**Assumption 1.** Input is persistently exciting of order $i + h$, and there is no feedback from output to input which means input signal and noise are uncorrelated.

**Assumption 2.** All transfer functions in (1) are bounded-input, bounded-output stable. This means that all poles lie inside the unit disk in complex plane.

**Assumption 3.** The maximum delay ($d_{max}$) is known.

## 3. Main Results

### 3.1. Recursive estimation of Markov Parameters

Consider a set of data generated by (2), and define a window of data of width $w$ as $D_w(k) \triangleq \{(u_{k-w+1}, y_{k-w+1}),\ \ (u_{k-w+2}, y_{k-w+2}),\ \ \dots,\ \ (u_k, y_k)\}$. Choose horizons $j$, $h$ and $i$ such that $w = j + i + h + 1$, $j \gg i, h$and $h > d_{max}$ to define Hankel matrices $U_p(k; w)$ and $U_f(k; w)$ according to (4) as

$$U_p(k; w) \triangleq U_{(k-j-h-i+2),i,j}\ U_f(k; w) \triangleq U_{(k-j-h+2),h,j} \qquad \textbf{(7)}$$

Hankel matrices $Y_p(k; w)$ and $Y_f(k; w)$ are defined similarly and the matrix containing the past input-output data is $W_p(k; w) \triangleq \begin{pmatrix} U_p(k; w) \\ Y_p(k; w) \end{pmatrix}$. Concatenated input/output vectors are created by putting several inputs/outputs into a column:

$$\begin{cases} u_s(k) \triangleq (u^T(k-s+1) \quad u^T(k-s+2) \quad \cdots \quad u^T(k))^T \\ y_s(k) \triangleq (y^T(k-s+1) \quad y^T(k-s+2) \quad \cdots \quad y^T(k))^T \\ w_s(k) \triangleq \begin{pmatrix} u_s(k) \\ y_s(k) \end{pmatrix} \end{cases} \qquad \textbf{(8)}$$

Now by performing the LQ factorization

$$\begin{pmatrix} U_f(k; k) \\ W_p(k; k) \\ Y_f(k, k) \end{pmatrix} = \begin{pmatrix} L_{11}(k) & 0 & 0 \\ L_{21}(k) & L_{22}(k) & 0 \\ L_{31}(k) & L_{32}(k) & L_{33}(k) \end{pmatrix} \begin{pmatrix} {Q_1}^T(k) \\ {Q_2}^T(k) \\ {Q_3}^T(k) \end{pmatrix} \qquad \textbf{(9)}$$

for $D_k(k)$,we can estimate $\Psi_h$ by means of (7). We now derive the needed formulas for our algorithm.

Suppose that the data couple $(u_{k+1}, y_{k+1})$ is available. According to (9), we have the following identity for the new data set$D_{k+1}(k+1)$:

$$\begin{pmatrix} U_f(k+1;k+1) \\ W_p(k+1;k+1) \\ Y_f(k+1;k+1) \end{pmatrix} = \begin{pmatrix} U_f(k;k) & u_h(k+1) \\ W_p(k;k) & w_i(k-h+1) \\ Y_f(k;k) & y_h(k+1) \end{pmatrix} \tag{10}$$

Let$\gamma$ be a positive number smaller than one and very close to it. We may approximate the left-hand side of (11) by

$$\begin{pmatrix} U_f(k+1;k+1) \\ W_p(k+1;k+1) \\ Y_f(k+1;k+1) \end{pmatrix} \approx \begin{pmatrix} \sqrt{\gamma}U_f(k;k) & u_h(k+1) \\ \sqrt{\gamma}W_p(k;k) & w_i(k-h+1) \\ \sqrt{\gamma}Y_f(k;k) & y_h(k+1) \end{pmatrix} \tag{11}$$

By post-multiplying both sides of (12) with their respective transposes, and equating the resulting terms we get (A different subspace identification method presented by [31] uses this mathematical idea in another context for fixed data size. It also does not consider the approximation in (12))

$$L_1(k+1) = \gamma L_1(k) + u_{\boldsymbol{h}}(k+1)u_{\boldsymbol{h}}^T(k+1) \tag{12}$$

$$L_2(k+1) = \gamma L_2(k) + w_i(k-\boldsymbol{h}+1)u_{\boldsymbol{h}}^T(k+1) \tag{13}$$

$$L_3(k+1) = \gamma L_3(k) + y_{\boldsymbol{h}}(k+1)u_{\boldsymbol{h}}^T(k+1) \tag{14}$$

$$L_4(k+1) = \gamma L_4(k) + w_i(k-\boldsymbol{h}+1)w_i^T(k-\boldsymbol{h}+1) \tag{15}$$

$$L_5(k+1) = \gamma L_5(k) + y_{\boldsymbol{h}}(k+1)w_i^T(k-\boldsymbol{h}+1) \tag{16}$$

where below definitions were used.

$$L_1(k) \triangleq U_f(k;k)U_f^T(k;k) = L_{11}(k)L_{11}^T(k) \tag{17}$$

$$L_2(k) \triangleq W_p(k;k)U_f^T(k;k) = L_{21}(k)L_{11}^T(k) \tag{18}$$

$$L_3(k) \triangleq Y_f(k,k)U_f^T(k;k) = L_{31}(k)L_{11}^T(k) \tag{19}$$

$$L_4(k) \triangleq W_p(k;k)W_p^T(k;k) = L_{21}(k)L_{21}^T(k) + L_{22}(k)L_{22}^T(k) \tag{20}$$

$$L_5(k) \triangleq Y_f(k,k)W_p^T(k;k) = L_{31}(k)L_{21}^T(k) + L_{32}(k)L_{22}^T(k) \tag{21}$$

Since the input is persistently exciting of order$i+h$ , $L_{11}(k)$ is full-rank (proof in [25]), and as a result $L_1(k)$ is invertible. Hence, from (18) and (20) we may write

$$\begin{aligned} L_{31}(k)L_{11}^{-1}(k) &= L_{31}(k)L_{11}^T(k)(L_{11}^T(k))^{-1}L_{11}^{-1}(k) \\ &= L_3(k)L_1^{-1}(k) \end{aligned} \tag{22}$$

Similarly, (18)-(22) yield

$$\begin{aligned} L_{32}(k)L_{22}^T(k) &= L_5(k) - L_{31}(k)L_{21}^T(k) \\ &= L_5(k) - L_{31}(k)L_{11}^T(k)(L_{11}^T(k))^{-1}L_{11}^{-1}(k)L_{11}(k)L_{21}^T(k) \\ &= L_5(k) - L_3(k)L_1^{-1}(k)L_2^T(k) \end{aligned} \tag{23}$$

$$\begin{aligned} L_{22}(k)L_{22}^T(k) &= L_4(k) - L_{21}(k)L_{21}^T(k) \\ &= L_4(k) - L_{21}(k)L_{11}^T(k)(L_{11}^T(k))^{-1}L_{11}^{-1}(k)L_{11}(k)L_{21}^T(k) \\ &= L_4(k) - L_2(k)L_1^{-1}(k)L_2^T(k) \end{aligned} \tag{24}$$

$$L_{21}(k)L_{11}^{-1}(k) = L_{21}(k)L_{11}^{T}(k)(L_{11}^{T}(k))^{-1}L_{11}^{-1}(k) \qquad \textbf{(25)}$$
$$= L_2(k)L_1^{-1}(k)$$

We now write (7) in terms of $L_i(k), i = 1,\dots,5$. For this, notice that the matrix equality $A^{\dagger} = A^T(AA^T)^{\dagger}$ holds for every matrix$A$. From this identity and equations (7) and (10), it can be written

$$\widehat{\boldsymbol{\Psi}}_{\boldsymbol{h}}(k) = L_{31}(k)L_{11}{}^{-1}(k) - (L_{32}(k)L_{22}{}^{T}(k))(L_{22}(k)L_{22}{}^{T}(k))^{\dagger}L_{21}(\boldsymbol{k})\,L_{11}{}^{-1}(k) \qquad \textbf{(26)}$$

Substitutions of (23)-(26) into (27) gives

$$\widehat{\boldsymbol{\Psi}}_{\boldsymbol{h}}(k) = \{L_3(k) - [L_5(k) - L_3(k)P(k)L_2{}^T(k)][L_4(k) - L_2(k)P(k)L_2{}^T(k)]^{\dagger}L_2(k)\}P(k) \qquad \textbf{(27)}$$

, in which$P(k) = L_1{}^{-1}(k)$.$P(k)$ can be calculated recursively using the matrix inversion lemma in order to reduce computational load:

$$P(k+1) = \frac{1}{\gamma}P(k)[I_1 - \frac{u_h(k+1)u_h{}^T(k+1)P(k)}{\gamma + u_h{}^T(k+1)P(k)u_h(k+1)}] \qquad \textbf{(28)}$$

Equations (13)-(17) along with (28) and (29) will be used for recursive calculation of Markov parameters. Before presenting the algorithm, a new criterion for time delay estimation should be introduced.

### *3.2. Choice of Criterion*

As pointed out earlier, comparison with a threshold for delay determination poses some problems, and one should use a different criterion instead. Consider the system in (1) and suppose that we are interested in estimating $T_{ji}$ whose true value is $d$. Further assume that $\widehat{\Psi}_h(k)$ is available. Take the first block column of this matrix (the first $m$ columns) and put the elements corresponding to the pair $(y_j, u_i)$ orderly into a vector $(\psi_k^{ji})$. The constructed vector will be of the following form.

$$\psi_k^{ji} = (\varepsilon_1 \quad \varepsilon_2 \quad \dots \quad \varepsilon_{d+1} \quad g_0 \quad g_1 \quad \dots)^T \qquad \textbf{(29)}$$

In this case, the first $d+1$ elements of $\psi_k^{ji}$ have very small magnitudes ($\varepsilon_i \ll 1, i = 1,2,\dots,d+1$). Therefore, the maximum absolute value of these entries is also very small. In addition, the absolute value of the $(d+2)$ th element $(g_0)$ has to be significantly larger than this maximum value, showing the first upsurge in impulse response. These facts lead us to a function of the form

$$C(t) = \left|\psi_k^{ji}(t+1)\right| / (\left\|\phi_k^{ji}(t+1)\right\|_{\infty} + \varepsilon) \qquad \textbf{(30)}$$

in which $\psi_k^{ji}(t+1)$ is the $(t+1)$ th element of $\psi_k^{ji}$, $\phi_k^{ji}(t) = \psi_k^{ji}(1{:}\,t-1)$ using MATLAB notation, and $\varepsilon$ is a very small positive number embedded to prevent the denominator from singularity. Now, the time delay can be determined by the following criterion.

$$d = arg\,\max_{t\geq 0} C(t) - 1 \qquad \textbf{(31)}$$

It is worth pointing out that the function in (31) automatically guarantees the fact that $\left|\psi_k^{ji}(d+2)\right|$ should be greater than the maximum value of the preceding elements.

Time delay estimation procedure can now be summarized as follows:

(I) **Choose** a large $N$ and assign suitable values to horizons$j$, $i$ and $h$. Note that $j$ must be large, and $h$ must be greater than $d_{max}$. Also, choose $\gamma$close to but less than unity.

(II) **Form** data matrices $U_f(N;N)$, $Y_f(N;N)$ and $W_p(N;N)$. Calculate $L_r(N)$,$r = 1,2,\ldots,5$, according to (18-22) and compute $P(N) = {L_1}^{-1}(N)$.

(III) **Use** these matrices to find $\widehat{\Psi}_h(N)$ by (28)**.**

(IV) **Determine** time delays by considering the first block column of $\widehat{\Psi}_h(N)$ and criterion (32).

(V) **Receive** the $(N+1)$ th data. Update $P(N+1)$ and$L_r(N+1)$,$r = 2,\ldots,5$, using the update formulas (14-17) and (29).

(VI) **Calculate** $\widehat{\Psi}_h(N+1)$ by means of (28), and find the new time delays' estimates according to (32).

(VII) **Increase** $N$ by one and go back to step (V).

**Remark 1.** Updating matrices by weighted sums of their old values and the terms containing new information $(u_{k+1}, y_{k+1})$ amounts to incorporation of forgetting factor. This way one can assign more weight to newer input-output data. From this perspective, $\gamma$ plays the role of forgetting factor in the update formulas (13-17). As mentioned, $\gamma$ is taken smaller than unity. Smaller choice of $\gamma$ apparently means that the weight assigned to new data is greater which is appealing. However, the approximation in (12) becomes less accurate which gives less accurate estimate of the Toeplitz matrix $\Psi_h(k+1)$. In addition, the accuracy of this approximation is dependent upon not only the choice of$\gamma$, but also the data itself which in turn depends on the system's intrinsic properties. Therefore, it is wiser to prioritize the accuracy of approximation (12) over the tracking speed of estimator by choosing $\gamma$ as close to unity as possible.

*3.3. Brief Discussion on Noise-related Issues*

Effectiveness of the presented algorithm mostly depends on the accuracy of the estimate in (7), which is guaranteed for noiseless data and large data size ($N$). Obviously, this estimate cannot be used for noisy data in general, but one may consider it for data corrupted by negligible noise. In fact, it is intuitive that (7) can be utilized for data with large signal-to-noise ratio. Nevertheless, we assume a large data size ($N \to \infty$) for derivation of subspace formulas like (7), and as a result, even a negligible noise may introduce a bias in our estimates. In such case, the effect of noise may be viewed as a summation of infinite number of very small numbers which can diverge or at least converge to a nonzero value. This problem therefore necessitates a theoretical investigation rather than relying solely on an intuitive approach. For this, consider the following MIMO state-space model

$$x(k+1) = Ax(k) + Bu(k)$$
$$y(k) = Cx(k) + Du(k) + e(k) \qquad \textbf{(32)}$$

where $e(k)$ is the additive noise and other notations are the same as those given for (2). For a window of data, the extended state-space form of this model is attained similar to (5) as

$$Y_p^s = \boldsymbol{\Gamma}_i X_p + \boldsymbol{\Psi}_i U_p + E_p$$
$$Y_f^s = \boldsymbol{\Gamma}_h X_f + \boldsymbol{\Psi}_h U_f + E_f \qquad \textbf{(33)}$$
$$X_f = A^i X_p + \boldsymbol{\Delta}_i U_p$$

where $E_p$ and $E_f$ are the past and future noise Hankel matrices respectively, and superscript "s" represents the stochastic nature. Deterministic and stochastic parts of the right-hand-sides of (35) are separate so that one may write

$$Y_p^s = Y_p^d + E_p$$

$$Y_f^s = Y_f^d + E_f$$
$$X_f = A^i X_p + \Delta_i U_p \tag{34}$$

with $Y_p^d = \Gamma_i X_p + \Psi_i U_p$, $Y_f^d = \Gamma_h X_f + \Psi_h U_f$ and "d" designating deterministic property. Arrange the quadruple $(Y_p^s, Y_f^s, U_p, U_f)$ in form of (6) and compute its LQ decomposition. Use (7) and (28) to define

$$\widehat{\Psi}_h(e,u) \triangleq \{L_3^s - [L_5^s - L_3^s P^s (L_2^s)^T][L_4^s - L_2^s P^s (L_2^s)^T]^\dagger L_2^s\} P^s \tag{35}$$

in which $L_a^s$ ,$a = 1,2,\ldots,5$, are calculated as in (18-22) for the aforementioned quadruple and $P^s = (L_1^s)^{-1}$. The following lemma holds:

**Theorem 1.** Suppose that the system in (34) is BIBO stable and that Assumption 1 holds. Further assume that noise $e$ in (34) is a zero mean white process, independent of the input signal with covariance$R_e(i,j) = E(e(i)e^T(j)) = R_e \delta_{ij}$, where $\delta_{ij}$ designates the Kronecker delta. Consider the data window $D_w(k)$ $(w = h + i + j - 1)$ with a large $w$ for which (37) is calculated with regard to (35), and define

$$\widehat{\Psi}_h(\bar{0},u) \triangleq \{L_3 - [L_5 - L_3 P {L_2}^T][L_4 - L_2 P {L_2}^T]^\dagger L_2\} P \tag{36}$$

where the matrices are acquired from the following relations holding for the same data set.

$$\begin{cases} P = (U_f {U_f}^T)^{-1} \\ L_2 = \begin{pmatrix} U_p \\ Y_p^d \end{pmatrix} {U_f}^T \\ L_3 = Y_f^d {U_f}^T \\ L_4 = \begin{pmatrix} U_p \\ Y_p^d \end{pmatrix} \begin{pmatrix} {U_p}^T & (Y_p^d)^T \end{pmatrix} \\ L_5 = Y_f^d \begin{pmatrix} {U_p}^T & (Y_p^d)^T \end{pmatrix} \end{cases}$$

Then, the subsequent holds for the Frobenius (or spectral) norm:

$$\forall \delta > 0 \quad \exists \varepsilon > 0 \quad \textbf{s.t. if} \quad \|R_e\| < \varepsilon \quad \rightarrow \quad \left\|\widehat{\boldsymbol{\Psi}}_{\boldsymbol{h}}(e,u) - \widehat{\boldsymbol{\Psi}}_{\boldsymbol{h}}(\bar{0},u)\right\| < \delta \tag{37}$$

**Proof.** According to (18-22) we have

$$(P^s)^{-1} = L_1^s = L_1 = U_f {U_f}^T \tag{38}$$

$$L_2^s = W_p^s {U_f}^T = (W_p^d + \begin{pmatrix} \bar{0} \\ E_p \end{pmatrix}) {U_f}^T = L_2 + \begin{pmatrix} \bar{0} \\ E_p {U_f}^T \end{pmatrix} \tag{39}$$

$$L_3^s = Y_f^s {U_f}^T = (Y_f^d + E_f) {U_f}^T = L_3 + E_f {U_f}^T \tag{40}$$

$$L_4^s = W_p^s (W_p^s)^T = (W_p^d + \begin{pmatrix} \bar{0} \\ E_p \end{pmatrix})((W_p^d)^T + \begin{pmatrix} \bar{0} & {E_p}^T \end{pmatrix}) {E_p}^T = L_4 + \begin{pmatrix} \bar{0} & W_p^d {E_p}^T \end{pmatrix} + \begin{pmatrix} \bar{0} \\ E_p (W_p^d)^T \end{pmatrix} + \begin{pmatrix} \bar{0} & \bar{0} \\ \bar{0} & E_p {E_p}^T \end{pmatrix} \tag{41}$$

$$L_5^s = Y_f^s (W_p^s)^T = (Y_f^d + E_f)((W_p^d)^T + \begin{pmatrix} \bar{0} & {E_p}^T \end{pmatrix}) = L_5 + \begin{pmatrix} \bar{0} & Y_f^d {E_p}^T \end{pmatrix} + E_f (W_p^d)^T + \begin{pmatrix} \bar{0} & E_f {E_p}^T \end{pmatrix} \tag{42}$$

where $W_p^s \triangleq \begin{pmatrix} U_p \\ Y_p^s \end{pmatrix}$ and $W_p^d \triangleq \begin{pmatrix} U_p \\ Y_p^d \end{pmatrix}$ are used. In (41) through (44) all zero matrices are of appropriate dimensions. Substitution of equations (40-44) into (37) gives

$$\widehat{\Psi}_h(e,u) = (L_3 + E_f {U_f}^T)(U_f {U_f}^T)^{-1} - \{L_5 - L_3 (U_f {U_f}^T)^{-1} {L_2}^T + \phi_1\}$$

$$\{L_4 - L_2(U_f{U_f}^T)^{-1}{L_2}^T + \phi_2\}^\dagger\ (L_2 + \begin{pmatrix} \bar{0} \\ E_p{U_f}^T \end{pmatrix})(U_f{U_f}^T)^{-1} \tag{43}$$

in which

$$\phi_1 = \begin{pmatrix} \bar{0} & Y_f^d{E_p}^T \end{pmatrix} + E_f(W_p^d)^T + \begin{pmatrix} \bar{0} & E_f{E_p}^T \end{pmatrix}\phi_1 - \begin{pmatrix} \bar{0} & E_f{U_f}^T(U_f{U_f}^T)^{-1}U_f{E_p}^T \end{pmatrix}$$
$$-\begin{pmatrix} \bar{0} & L_3(U_f{U_f}^T)^{-1}U_f{E_p}^T \end{pmatrix} - E_f{U_f}^T(U_f{U_f}^T)^{-1}{L_2}^T \tag{44}$$

and

$$\phi_2 = \begin{pmatrix} \bar{0} & W_p^d{E_p}^T \end{pmatrix} + \begin{pmatrix} \bar{0} \\ E_p(W_p^d)^T \end{pmatrix} + \begin{pmatrix} \bar{0} & \bar{0} \\ \bar{0} & E_p{E_p}^T \end{pmatrix} - \begin{pmatrix} \bar{0} & L_2(U_f{U_f}^T)^{-1}U_f{E_p}^T \end{pmatrix}$$
$$-\begin{pmatrix} \bar{0} \\ E_p{U_f}^T(U_f{U_f}^T)^{-1}{L_2}^T \end{pmatrix} - \begin{pmatrix} \bar{0} & \bar{0} \\ \bar{0} & E_p{U_f}^T(U_f{U_f}^T)^{-1}U_f{E_p}^T \end{pmatrix} \tag{45}$$

Equation (45) can be written

$$\widehat{\Psi}_h(e,u) = (L_3 + E_f{U_f}^T)(U_f{U_f}^T)^{-1} - \{\frac{L_5 - L_3(U_f{U_f}^T)^{-1}{L_2}^T}{j} + \frac{\phi_1}{j}\}$$
$$\{\frac{L_4 - L_2(U_f{U_f}^T)^{-1}{L_2}^T}{j} + \frac{\phi_2}{j}\}^\dagger\ (L_2 + \begin{pmatrix} \bar{0} \\ E_p{U_f}^T \end{pmatrix})(U_f{U_f}^T)^{-1} \tag{46}$$

It is straightforward to show that $E_p{U_f}^T(U_f{U_f}^T)^{-1}$, $E_f{U_f}^T(U_f{U_f}^T)^{-1} \to \bar{0}$ as $j \to \infty$ since $u$ and $e$ are independent. Similarly, it can be shown that

$$\frac{1}{j}\phi_1 \to \bar{0} \quad \text{as} \quad j \to \infty \tag{47}$$

and

$$\frac{1}{j}\phi_2 \to \begin{pmatrix} \bar{0} & \bar{0} \\ \bar{0} & \frac{1}{j}E_p{E_p}^T \end{pmatrix} \quad \text{as} \quad j \to \infty \tag{48}$$

, where the latter leads to

$$\frac{1}{j}\phi_2 \to \begin{pmatrix} \bar{0} & \bar{0} \\ \bar{0} & I_{i\times i} \otimes R_e \end{pmatrix} \quad \text{as} \quad j \to \infty \tag{49}$$

By defining $\Lambda_1 \triangleq \frac{L_5 - L_3(U_f{U_f}^T)^{-1}{L_2}^T}{j}$ and $\Lambda_2 \triangleq \frac{L_4 - L_2(U_f{U_f}^T)^{-1}{L_2}^T}{j}$, (48) becomes

$$\widehat{\boldsymbol{\Psi}}_{\boldsymbol{h}}(e,u) = L_3(U_f{U_f}^T)^{-1} - \boldsymbol{\Lambda}_1\{\boldsymbol{\Lambda}_2 + \boldsymbol{\Upsilon}\}^\dagger L_2(U_f{U_f}^T)^{-1} \tag{50}$$

in which $\Upsilon = \begin{pmatrix} \bar{0} & \bar{0} \\ \bar{0} & I_{i\times i} \otimes R_e \end{pmatrix}$. It is easy to show that $\lim_{j\to\infty}\Lambda_1$ and $\lim_{j\to\infty}\Lambda_2$ exist for a BIBO stable system. We prove the lemma for the Frobenius norm (the case for spectral norm is proved by following the same procedure). The Frobenius norm of the difference between $\widehat{\Psi}_h(e,u)$ and $\widehat{\Psi}_h(\bar{0},u)$ is

$$\left\|\widehat{\boldsymbol{\Psi}}_{\boldsymbol{h}}(e,u) - \widehat{\boldsymbol{\Psi}}_{\boldsymbol{h}}(\bar{0},u)\right\|_F = \left\|\boldsymbol{\Lambda}_1\{(\boldsymbol{\Lambda}_2 + \boldsymbol{\Upsilon})^\dagger - {\boldsymbol{\Lambda}_2}^\dagger\}L_2(U_f{U_f}^T)^{-1}\right\|_F \tag{51}$$

Assume that $\alpha$ is a positive real number smaller than 1 and that we have

$$\begin{cases} \|R_e\|_F \le \boldsymbol{min}(\alpha\sigma 2\frac{\delta}{\boldsymbol{\Xi_{min}}} \\ \boldsymbol{\Xi} = \dfrac{\sqrt{i}}{(1-\alpha)\sigma 2_{2_{\boldsymbol{min}}}\|\boldsymbol{\Lambda_1}\|_F \left\|\frac{L_2}{j}\right\|_F \left\|(U_f U_f^{\ T})^{-1}\right\|_F} \{ \end{cases} \quad \textbf{(52)}$$

where $\sigma 2_{min}$ is the smallest singular value of $\Lambda_2$. Since$\|R_e\|_F \le \alpha\sigma 2_{min}$, we get

$$\|R_e\|_2 \le \|R_e\|_F \le \alpha\sigma 22_{\boldsymbol{min}_{\boldsymbol{min}}}$$
$$\Rightarrow \quad \|I_{i\times i} \otimes R_e\|_2 < \sigma 2_{min} \quad \textbf{(53)}$$

This implies that ranks of matrices $\Lambda_2$ and $(\Lambda_2 + \Upsilon)$ are equal. Consider Lemma 1 (or Lemma 2 for the spectral norm) to achieve an upper bound for $\left\|(\Lambda_2 + \Upsilon)^\dagger - {\Lambda_2}^\dagger\right\|_F$ as

$$\begin{aligned} \left\|(\Lambda_2 + \Upsilon)^\dagger - {\Lambda_2}^\dagger\right\|_F &\le \left\|{\Lambda_2}^\dagger\right\|_2 \left\|(\Lambda_2 + \Upsilon)^\dagger\right\|_2 \left\|I_{i\times i} \otimes R_e\right\|_F \\ &\le \frac{1}{\sigma_{\min}(\Lambda_2)\sigma_{\min}(\Lambda_2 + \Upsilon)} \left\|I_{i\times i} \otimes R_e\right\|_F \\ &\le \frac{1}{\sigma_{\min}(\Lambda_2)(\sigma_{\min}(\Lambda_2) - \left\|I_{i\times i} \otimes R_e\right\|_2)} \left\|I_{i\times i} \otimes R_e\right\|_F \\ &\le \frac{1}{{\sigma_{\min}}^2(\Lambda_2)(1-\alpha)} \left\|I_{i\times i} \otimes R_e\right\|_F \end{aligned}$$

**(54)**

Hence, (53) becomes

$$\begin{aligned} \left\|\hat{\Psi}_h(e,u) - \hat{\Psi}_h(\bar{0},u)\right\|_F &= \left\|\Lambda_1\{(\Lambda_2 + \Upsilon)^\dagger - {\Lambda_2}^\dagger\} L_2 (U_f {U_f}^T)^{-1}\right\|_F \\ &\le \left\|\Lambda_1\right\|_F \left\|\{(\Lambda_2 + \Upsilon)^\dagger - {\Lambda_2}^\dagger\}\right\|_F \left\|L_2\right\|_F \left\|(U_f {U_f}^T)^{-1}\right\|_F \\ &\le \left\|\Lambda_1\right\|_F \frac{1}{{\sigma_{\min}}^2(\Lambda_2)(1-\alpha)} \left\|I_{i\times i} \otimes R_e\right\|_F \left\|L_2\right\|_F \left\|(U_f {U_f}^T)^{-1}\right\|_F \\ &= \left\|\Lambda_1\right\|_F \frac{\sqrt{i}}{{\sigma_{\min}}^2(\Lambda_2)(1-\alpha)} \left\|R_e\right\|_F \left\|\frac{L_2}{j}\right\|_F \left\|(\frac{U_f {U_f}^T}{j})^{-1}\right\|_F \end{aligned} \quad \textbf{(55)}$$

(54) and (57) prove the lemma:

$$\begin{aligned} \left\|\hat{\Psi}_h(e,u) - \hat{\Psi}_h(\bar{0},u)\right\|_F &\le \left\|\Lambda_1\right\|_F \left\|\frac{L_2}{j}\right\|_F \frac{\sqrt{i}}{{\sigma_{\min}}^2(\Lambda_2)(1-\alpha)} \left\|R_e\right\|_F \left\|(\frac{U_f {U_f}^T}{j})^{-1}\right\|_F \\ &\le \left\|\Lambda_1\right\|_F \left\|\frac{L_2}{j}\right\|_F \frac{\sqrt{i}}{{\sigma_{\min}}^2(\Lambda_2)(1-\alpha)} \left\|(\frac{U_f {U_f}^T}{j})^{-1}\right\|_F \frac{\delta}{\Xi} = \delta \end{aligned} \quad \textbf{(56)}$$

□

**Remark 2.** This Lemma states that the presented algorithm has good performance for white and sufficiently small measurement noise which is independent of input excitation. It can be shown that the converse is also true as expected, i.e. for any noise covariance matrix, a proper excitation can give accurate results. For instance, suppose that the input to a special SISO system is a zero-mean sequence of independent random (or pseudo-random) variables denoted by $\mathfrak{I} = \{u(1), u(2), \dots, u(j)\}$. For this sequence, $\Xi$in inequality (54) becomes

$$\Xi_1 = \frac{\sqrt{i}}{(1-\alpha)\sigma 2_{2_{min}} \|\Lambda_1\|_F \left\| \frac{L_2}{j} \right\|_F \|(I_{h\times h} \otimes R_u)^{-1}\|_F \dfrac{\sqrt{ih}}{R_u (1-\alpha)\sigma 2_{2_{min}} \|\Lambda_1\|_F \left\| \frac{L_2}{j} \right\|_F}} \quad \textbf{(57)}$$

If we apply a new input sequence $\mathfrak{I}' = \beta\mathfrak{I} = \{\beta \boldsymbol{u}(\mathbf{1}), \beta \boldsymbol{u}(\mathbf{2}), \dots, \beta \boldsymbol{u}(\boldsymbol{j})\}$ with $\beta > 1$ to the system, each of$\Lambda_1$, $\Lambda_2$, $L_2$, $R_u$and $\sigma 2_{min}$ will be multiplied by $\beta^2$. Consequently, we get for the second sequence

$$\Xi_2 = \frac{\sqrt{ih}}{\beta^2 R_u (1-\alpha)\sigma 2_{2_{min}} \|\Lambda_1\|_F \left\| \frac{L_2}{j} \right\|_F \frac{1}{\beta^2}_1} \quad \textbf{(58)}$$

According to lemma 1, the following must be satisfied to ensure the $\delta$ -accuracy in (39):

For $\mathfrak{I}$: $R_e \leq min(\alpha\sigma 2 \frac{\delta}{\Xi_{1_{min}}}$

and

For $\mathfrak{I}'$: $R_e \leq \beta^2\, min(\alpha\sigma 2 \frac{\delta}{\Xi_{1_{min}}}$

This indicates that for every$R_e$, one can find an input sequence (or equivalently a $\beta$ for any specific sequence) so that $\|\hat{\Psi}_h(e,u) - \hat{\Psi}_h(\bar{0},u)\| < \delta$, for every $\delta$. In particular, if $R_e$ becomes $\beta R_e$, the sequence $\mathfrak{I}$ should be multiplied by $\sqrt{\beta}$, or equivalently $R_u$ must become $\beta R_e$. Nevertheless, $min(\alpha\sigma 2 \frac{\delta}{\Xi_{1_{min}}}$ depends on both the input sequence and the model and cannot be determined if no prior information is accessible. In such situations, it is advisable to take $h$ as large as possible, and excite the system by the largest allowable input signal to ensure the optimal performance of the presented algorithm. It should be mentioned that the above conclusion is only valid for additive white noise, and no such a guarantee exists for the case of colored noise however small it would be.

**Remark 3.** *Application to SARX models:* consider a Switched Auto Regressive eXogenous (SARX) model with input-delay which can be described bys

$$y_t = \sum_{j=1}^{n_a(\lambda_t)} a_j(\lambda_t) y_{t-j} + \sum_{k=1}^{n_b(\lambda_t)} b_k(\lambda_t) u_{t-k-d(\lambda_t)} + e_t \quad \textbf{(59)}$$

in which$u_t \in \mathbb{R}$is the input, $y_t \in \mathbb{R}$is the output, $e$ is white noise and $\lambda_t = \{1,2,\dots,s\}$is the discrete state or equivalently *mode* of the system. The submodel corresponding to state $\lambda_t$ is characterized by the coefficients $a_j(\lambda_t)$ and $b_k(\lambda_t)$, where$j = 1,\dots,n_a(\lambda_t)$ and $k = 1,\dots,n_b(\lambda_t)$, and the time delay $d(\lambda_t)$. Transitions between the submodels are governed by a switching mechanism which is usually unknown. To the best of our knowledge, the majority of the available identification methods for SARX models consider delay-free model structures which are in general not suitable for handling systems with time delays. The problem is better noticed when the approach being used presumes equal orders for submodels (see [32], [33], [34]). In such cases, identification may lead to delay-free structures of very high orders due to existence of delayed submodels as in (60). It is therefore necessary to have delay estimates in order to take them into account when performing any identification algorithm. Since we normally do not know about the mode sequence, nor do we have information on parameter vectors (i.e. coefficients in (60)), time delay estimation should be done in a black-box fashion. This suggests that the method developed in this paper can be a candidate for estimation of time delays in (60) provided that $e$ is small enough. We should also restrict our target systems to cases of sufficiently small switching frequencies for successful time delay estimation. It should be noted that high switching frequencies can lead to very high order submodels due to varying nature of $d(\lambda_t)$. In addition, the maximum value of delay should be known (Assumption 3) since delay estimates will be otherwise smaller than their true magnitudes.

## 4. Numerical Results

Example 1. Consider the following multivariable model with two inputs and two outputs.

$$\begin{cases} y_1(k) = \dfrac{0.02q^{-1} + 0.1q^{-2}}{1 - 1.1q^{-1} + 0.3q^{-2}} u_1(k - 1 - T_{11}(k)) \\ + \dfrac{q^{-1} + 2q^{-2}}{1 - 0.4q^{-1}} u_2(k - 1 - T_{12}(k)) + e_1(k) \\ y_2(k) = \dfrac{q^{-1} + 0.5q^{-2}}{1 + 0.9q^{-1}} u_1(k - 1 - T_{21}(k)) \\ + \dfrac{0.02q^{-1}}{1 - 0.8q^{-1} - 0.25q^{-2} + 0.2q^{-3}} u_2(k - 1 - T_{22}(k)) \\ + e_2(k) \end{cases} \tag{61}$$

$e_1$ and $e_2$ are independent white Gaussian noises with equal standard deviations of 0.01, and delays' values over an interval of 2000 samples are specified below.

$$T_{11} = \begin{cases} 6 & 1 \le k \le 500 \\ 2 & 500 < k \le 2000 \end{cases} \quad T_{12} = \begin{cases} 1 & 1 \le k \le 500 \\ 5 & 500 < k \le 1000 \\ 2 & 1000 < k \le 2000 \end{cases}$$

$$T_{21} = \begin{cases} 1 & 1 \le k \le 1500 \\ 4 & 1500 < k \le 2000 \end{cases} \quad T_{22} = \begin{cases} 5 & 1 \le k \le 500 \\ 2 & 500 < k \le 1500 \\ 1 & 1500 < k \le 2000 \end{cases}$$

The system is excited by two pseudo-random binary sequences (PRBS) of frequency $\frac{1}{2047(\text{sample})}$ each, and the input-output data is available. The maximum value of delays is considered to be known. Accordingly, we choose the following parameters in applying the proposed delay estimation algorithm:

$j = 100, i = 8, h = 8, N = 115, \gamma = 0.9$ and $\varepsilon = 10^{-5}$

Fig.1 shows the resulted estimates, and as it is seen, the time delays are successfully estimated.

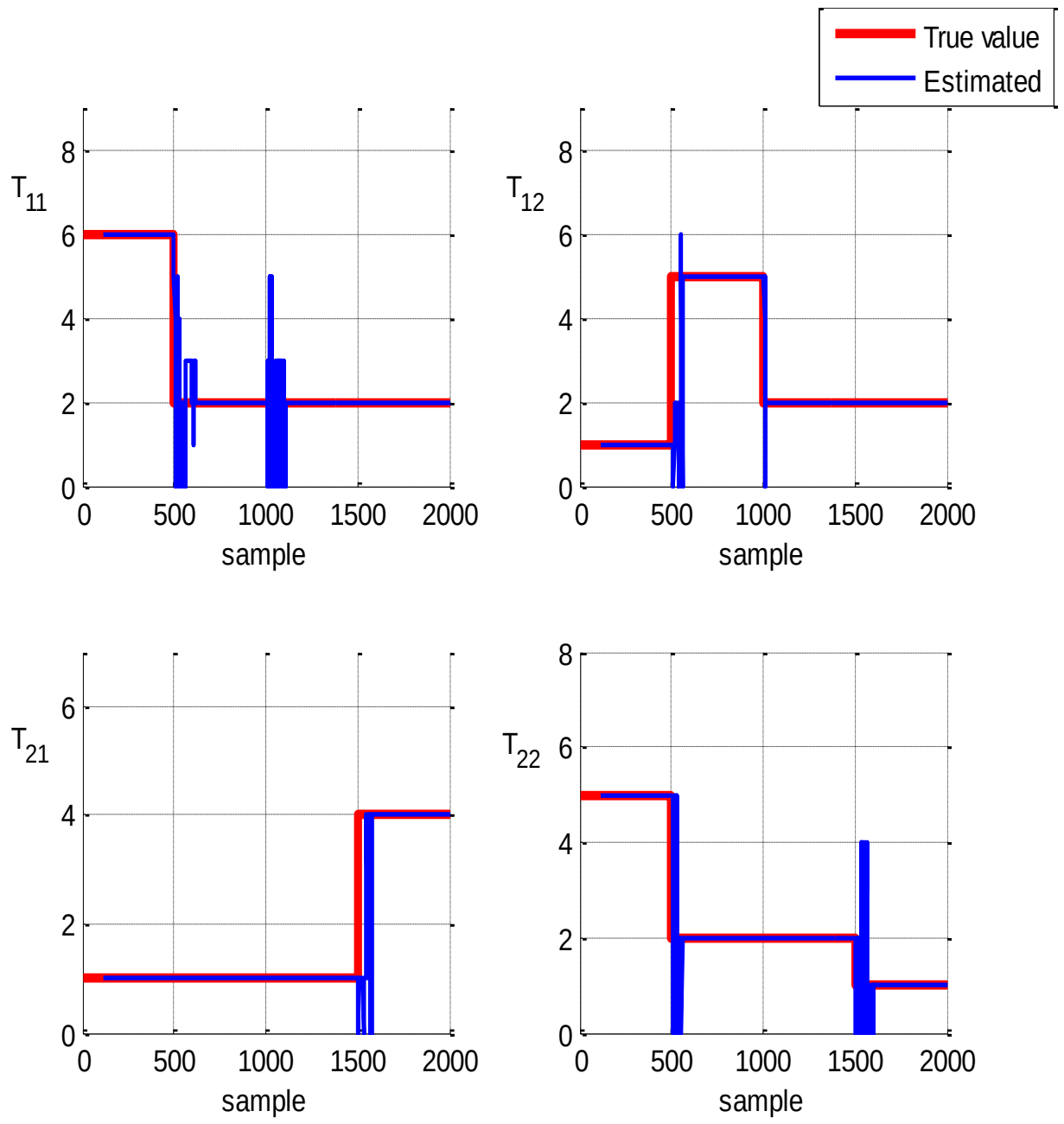

**Fig 1. Delay estimates (blue lines) using the proposed algorithm**

According to Fig.1, a change in the value of $T_{12}$at $k = 1000$ has affected $T_{11}$ estimates around this sample. A similar change has occurred for $T_{22}$ at $k = 500$where it went from 5 to 2. However, $T_{21}$estimates have not been disturbed around$k = 500$. Therefore, Fig.1 suggests that changes in a time delay may or may not affect estimation of other delays in the same loop.

As pointed out earlier, the criterion proposed by [28] fails to give consistent estimates when Markov parameters are very small. To illustrate this, we repeat the algorithm using this criterion instead of (32). The resulted estimates are shown in Fig.2. Notice that estimates of $T_{11}$ and $T_{22}$ are generally larger than their true values, showing the ineffectiveness of the used criterion.

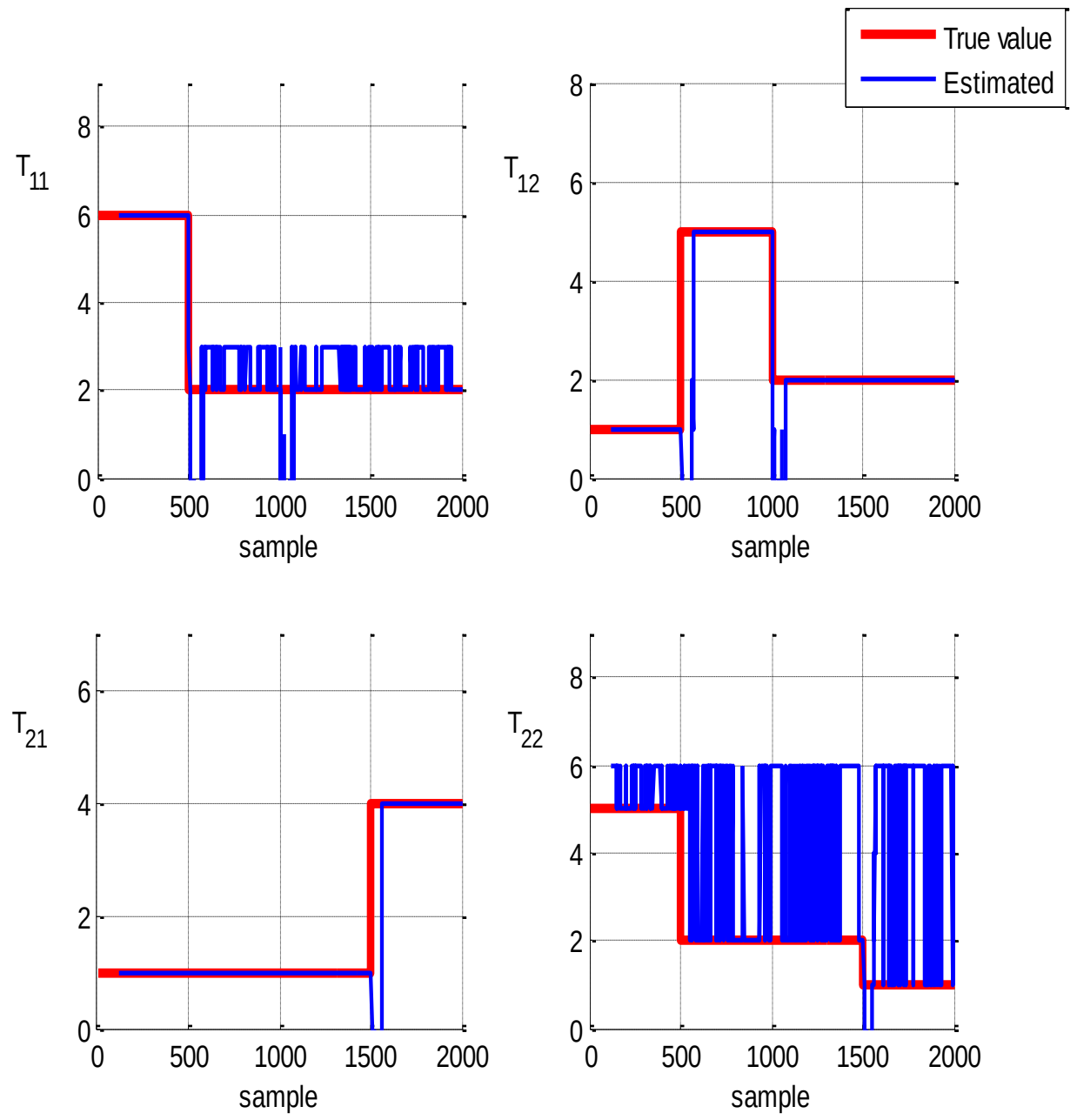

**Fig 2. Computed estimates using the criterion proposed by [16]**

**Example 2.** Consider the delayed SARX model

$$y(t) = \theta_{\lambda_t}^T \varphi_{\lambda_t}(t) + e(t) \tag{60}$$

where $\lambda_t \in \{1,2,3\}$. Respective parameters and regressors for each mode is given below.

$$\begin{cases} \theta_1 = (0.89 \quad -0.0710 \quad -1.2878 \quad -1.1252)^T \\ \varphi_1(t) = (y(t-1) \quad y(t-3) \quad u(t-2) \quad u(t-3))^T \\ \theta_2 = (-0.75 \quad 1.1050 \quad 3.16)^T \\ \varphi_2(t) = (y(t-1) \quad u(t-4) \quad u(t-5))^T \\ \theta_3 = (-0.5 \quad 0.1875 \quad 0.4055)^T \\ \varphi_3(t) = (y(t-1) \quad y(t-2) \quad u(t-5))^T \end{cases} \tag{61}$$

Suppose that $e$ is a white Gaussian noise with variance$\sigma^2$, and that switching between submodels occur every 250 samples. For delay estimation, the system is excited by a zero-mean Gaussian sequence of length 2000 with unit variance. Consider a random switching evolution of(3,2,3,1,2,3,1,2). Time delay $d$ is also depicted in Fig.3 according to (47). We have applied the algorithm to the generated data using$j = 78$,$i = 11$,$h = 12$,$N = 100$,$\gamma = 0.8$,$\varepsilon = 10^{-5}$ and assuming a standard deviation of 0.05 for noise which gives a signal-to-noise ratio of 20 dB . The estimated delay is depicted in Fig.4. As it is seen, delays are successfully estimated after each switching.

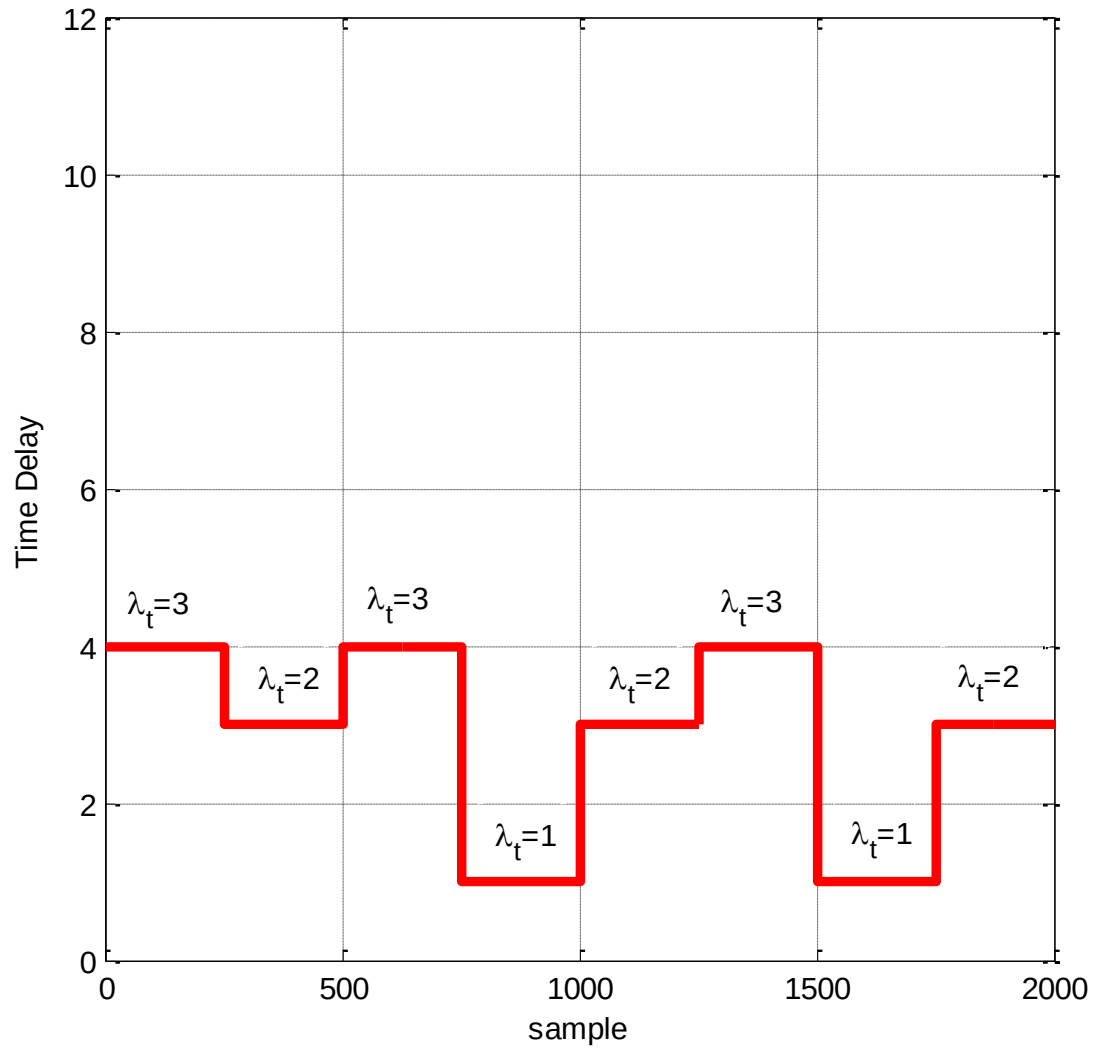

**Fig 3. Time delay evolution of model (62) with switching frequency of $\frac{1}{250}(1/\text{sample})$.**

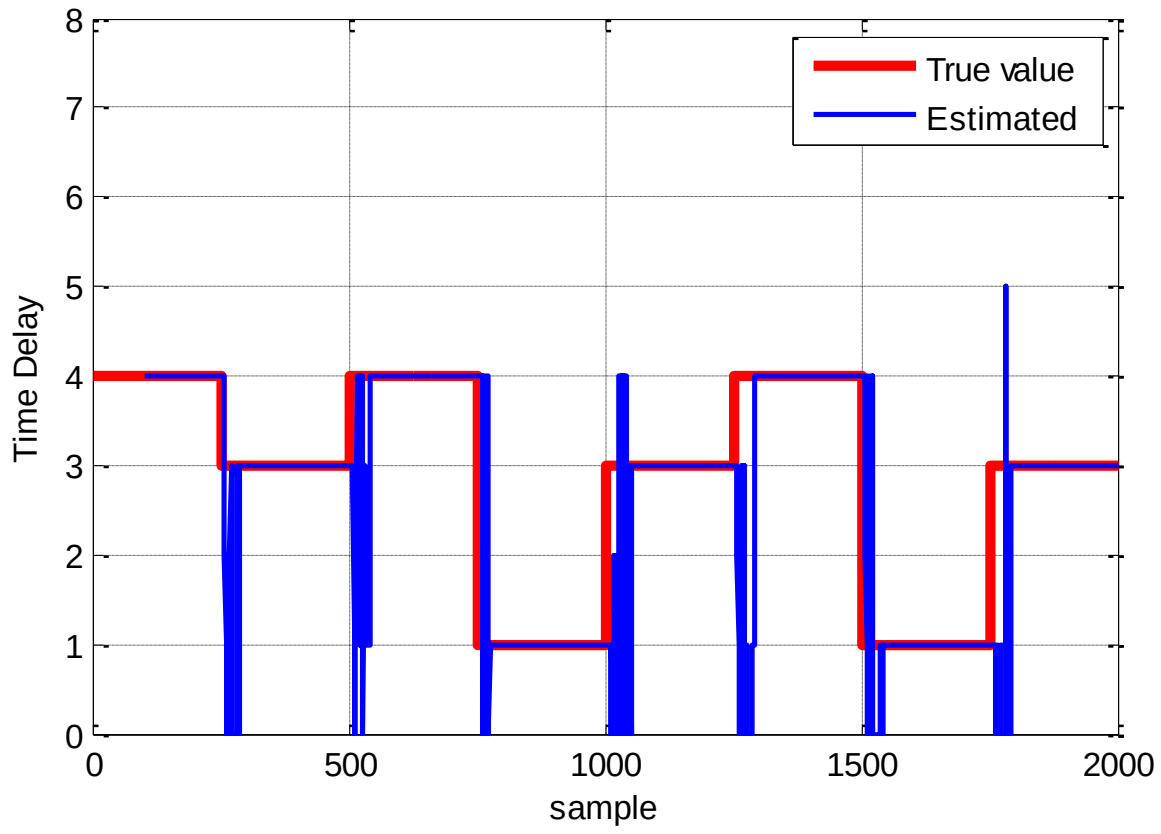

**Fig 4. Time delay estimates are shown by the blue lines tracking the true values (red lines) after each switching.**

## 5. CONCLUSIONS

An algorithm was developed to estimate delays in linear multiple-input multiple-output models with input delays. It recursively computes Markov parameters by which time delays are determined according to the proposed criterion. Forgetting factor was incorporated into the algorithm using an approximation in derivation of update formulas. The new criterion, in contrast to the one suggested by previous studies, also works well for the case of small Markov parameters. It uses the infinity norm which does not add significant computational load, making the algorithm easily implementable. The algorithm shows good performance when the additive noise is small. However, the needed excitation cannot be determined without extra a priori knowledge. Developed in subspace framework, the algorithm is appropriate for situations in which black-box estimation/identification is inevitable. This merit makes the presented algorithm applicable to delayed switched ARX models with the assumption that switching frequency is sufficiently small. Otherwise, the problem of estimating delays may be ill-posed. In other words, time delay effect may not be “distinguishable” from that of delay-free dynamics, in which case higher order submodels will be achieved.

Future work will focus primarily on robustification of the algorithm against measurement or other sources of noise. Alternatively, one may apply the same procedure used in deriving this paper’s results to other subspace methods that manifest better performance in dealing with noise. Of course, choice of such methods is not trivial as candidate methods should be structurally suitable for the procedure to be applicable. Considering noisy data, performance of the proposed algorithm degrades if magnitudes of Markov parameters are very small. Addressing this problem (by explicit delay estimation for example), is also desirable in future studies.